\documentclass[twocolumn,prl,aps,superscriptaddress]{revtex4-2}

\usepackage{amssymb,amsmath}
\usepackage{graphicx}
\usepackage{amsmath}		
\usepackage{setspace}
\usepackage{color}
\usepackage{collcell}
\usepackage{datatool}
\usepackage{environ}
\usepackage{latexsym}
\usepackage{amssymb}
\usepackage{epsfig,amsmath,graphics}
\usepackage{epstopdf}
\usepackage{verbatim}
\usepackage{wasysym}
\usepackage{feynmp-auto}
\usepackage{xcolor}
\usepackage{enumitem}
\usepackage[utf8]{inputenc}
\usepackage{slashed}
\usepackage[toc,page]{appendix}
\usepackage[normalem]{ulem}

\usepackage{empheq}

    \newlength\fsep
    \newsavebox\widebox

\definecolor{deepgreen}{rgb}{0.2,0.8,0.3}

\definecolor{deepblue}{rgb}{0.2,0.2,0.8}

\definecolor{deepred}{rgb}{0.8,0.2,0.2}

\newcommand{\dbraket}[1]{\langle\!\langle #1 \rangle\!\rangle}

\begin{document}
\title{A Solution to the Hierarchy Problem with Non-Linear Quantum Mechanics}

\author{David E.~Kaplan}
\affiliation{Department of Physics \& Astronomy, The Johns Hopkins University, Baltimore, MD  21218, USA}
\affiliation{Kavli IPMU (WPI), UTIAS, The University of Tokyo, Kashiwa, Chiba 277-8583, Japan}

\author{Surjeet Rajendran}
\affiliation{Department of Physics \& Astronomy, The Johns Hopkins University, Baltimore, MD  21218, USA}

\date{\today}

\begin{abstract}
We argue that the hierarchy problem of the standard model of particle physics can be solved by adding a state-dependent term to the Higgs sector.  We present an example of a scalar field with a Higgs-like potential with an additional term proportional to the expectation value of the squared Higgs field operator.  We show that the mass can be parametrically lighter than the theory's energy-momentum cutoff without fine tuning.  We find the Higgs mass can be technically natural, even with a Planck-scale cutoff.  The simplest version of the theory may not be distinguishable from the standard model at colliders, but  other versions might.  In addition, some aspects of cosmological evolution can be different in this model, in some cases radically.
\end{abstract}

\maketitle




The smallness of the Higgs mass in the Standard Model with respect to assumed higher scale physics is a mystery \footnote{A first discussion appears in print in \cite{PhysRevD.20.2619}, but Susskind (and many others) credit Ken Wilson for the original insight of a naturalness problem.  It is more thoroughly discussed in 't Hooft's Cargese lectures \cite{tHooft:1980xss}.}.  A natural explanation is a cutoff of the Standard Model at around the Higgs mass where new degrees of freedom should show up.  In these theories, the Higgs mass is effectively proportional to a different dimensionful scale (say of symmetry breaking \cite{Dimopoulos:1981zb}, compositeness \cite{Kaplan:1983fs}, or quantum gravity \cite{Arkani-Hamed:1998jmv}) which is stable relative to quantum corrections.  There is a growing tension between these types of solutions and experimental data, due to both collider constraints \cite{ParticleDataGroup:2024cfk} and indirect probes, such as flavor physics or CP violation (see, for example, \cite{Glioti:2024hye}).

Without symmetries or dynamics that make the global vacuum of the theory special, there is a Weinberg `no-go' theorem (originally applied to the cosmological constant problem \cite{Weinberg:1988cp}) suggesting that the mass of a fundamental Higgs particle must be fine-tuned.  In order to make the Higgs mass small at the minimum of its potential $V(\phi)$, its first derivative and second derivative must approximately vanish at the same point: $dV/d\phi = 0$ and $d^2V/d\phi^2 \simeq 0$ at the same value of $\phi$.  This problem is solved by having many minima (many solutions to $dV/d\phi = 0$), and then giving a dynamical \cite{Graham:2015cka} or anthropic \cite{Arkani-Hamed:2004ymt} reason that we exist in the minimum with $d^2V/d\phi^2 \simeq 0$.  (Other solutions involving  cosmological dynamics include theories with many sectors \cite{Dvali:2007hz,Arkani-Hamed:2016rle}).

Here we present a novel way to solve the hierarchy problem without new symmetries while addressing Weinberg's no-go \footnote{The basic idea discussed here can also potentially be applied to get around the Weinberg no-go theorem for the cosmological constant. We leave this for future work.}. It involves invoking non-linear quantum mechanics \cite{Kibble:1978vm,Kaplan:2021qpv}, namely adding state-dependent terms in the Hamiltonian in the form of expectation values of bosonic operators (see also \cite{Kibble:1978tm}).  Adding such terms allows one to dramatically manipulate the Higgs potential without adding new quantum effects.  Thus, despite being a more exotic form of a quantum field theory, these terms in fact make the theory more classical and thus offer a new way to control quantum effects.

By adding these terms to the Hamiltonian, one gives up manifest Lorentz invariance, as the theory is defined in a particular frame.  That said, by choosing couplings appropriately, the theory retains a Lorentz-invariant vacuum and, in that sense, Lorentz invariance is natural.






First let's discuss the standard hierarchy problem in the Hamiltonian language.  We do it in a bit of a pedantic way, noting the state of minimum energy that we are interested in will be approximately a coherent state.  We do this to prepare for the addition of state-dependent terms.

Let us start with a simple model of a complex scalar field, $\phi$, with a Hamiltonian
\begin{equation}\label{eq:LHamiltonian}
    \hat H_0 = \int d^3x \:\left(\hat\pi^\dagger\hat\pi + \nabla\hat\phi^\dagger \nabla\hat\phi - m_b^2 \hat\phi^\dagger\hat\phi + \frac{\lambda}{4}(\hat\phi^\dagger\hat\phi)^2\right)
\end{equation}
in the Schr\"{o}dinger picture with the equal-time commutator $\bigl[\pi(x),\phi(y)\bigr]=i\delta(x-y)$.  From the Schr\"{o}dinger equation, one can compute the time evolution of the operator expectation value $\langle\hat\phi\rangle$:
\begin{eqnarray}\label{eq:LinEOM}
    \partial_t \langle \hat\phi \rangle &=& i \langle\bigl[ \hat H_0 , \hat\phi \bigr]\rangle = \langle \hat\pi^\dagger \rangle \nonumber\\ \label{eq:LQMeom}
    \partial^2_t\langle\hat\phi\rangle  &=& \partial_t\langle\hat\pi^\dagger\rangle =  i\langle \bigl[\hat H_0 , \hat\pi^\dagger \bigr]\rangle \\ &=& \nabla^2\langle\hat\phi\rangle + m_b^2 \langle\hat\phi\rangle - \frac{\lambda}{2}\langle(\hat\phi^\dagger\hat\phi) \,\hat\phi\rangle \nonumber
\end{eqnarray}
This equation is exact, and one might think that one can simply set the bare mass $m_b^2$ to the required value without tuning.  However, the final term is the culprit.  Rewriting $\phi = (\varphi + i\chi)/\sqrt{2}$, one can see that, for example 
\begin{equation}
    \langle\hat\varphi(\hat\varphi^2 + \hat\chi^2)\rangle = \langle\varphi\rangle(\langle\hat\varphi\rangle^2 + \langle\hat\chi\rangle^2) + \Lambda^2\langle\hat\varphi\rangle + \cdots
\end{equation}
where $\Lambda$ should roughly be the (U(1)-invariant) energy-momentum cutoff. We have removed the `zero-point energy' and the remaining expectation values are regulated (cutoff-independent) \footnote{The remaining terms include wave-function renormalization and finite state dependent terms which will not affects the hierarchy problem in a qualitative way.}.  Thus, the equation of motion for $\varphi$ has an effective squared mass term, $m_r^2 \simeq m_b^2 + \Lambda^2$ and requires fine-tuning if $m_r^2 \ll \Lambda^2$.  Additional couplings to the scalar (say to fermions or gauge fields) will contribute in similar ways.

Consider now a static, homogeneous solution to \eqref{eq:LQMeom}:
\begin{equation}\label{eq:Leom-homo}
    0  =  m_r^2 \langle\hat\varphi\rangle - \frac{\lambda}{4}(\langle\hat\varphi\rangle^3 + \langle\hat\varphi\rangle \langle\hat\chi\rangle^2)
\end{equation}
From here we will take $m_r^2 > 0$. Assuming a narrow, coherent state, and a state where $\langle\hat\chi\rangle=0$, we have solutions $\langle\hat\varphi\rangle=0$ and $\varphi_0^2\equiv\langle\hat\varphi\rangle^2 = m_r^2/(\lambda/4)$.  Around the trivial solution, the excitations are unstable, while around $\langle\hat\varphi\rangle=m_r^2/(\lambda/4)$, the masses of particle excitations can be read off of \eqref{eq:LHamiltonian} by expanding the operator $\hat\varphi=\varphi_0 + \delta\hat\varphi$ and looking at the terms quadratic in $ \delta\hat\varphi$ and $\hat\chi$.  The latter is massless (the Nambu-Goldstone boson), while the former's squared mass can be found by taking a derivative of \eqref{eq:Leom-homo} with respect to $\langle\hat\varphi\rangle$ and inserting the vacuum expectation values (vevs):
\begin{eqnarray}
    \partial_t^2\langle\delta{\hat\varphi}\rangle  =  m_r^2\langle\delta\hat\varphi\rangle  - \frac{3\lambda}{4}\varphi_0^2\langle\delta\hat\varphi\rangle = -\frac{\lambda}{2}\varphi_0^2\langle\delta\hat\varphi\rangle
\end{eqnarray}
The physical squared mass of a particle created in this case is $m_{phys}^2 = (\lambda/2)\varphi_0^2$.  The squared masses and vacuum expectation values (vevs) of the scalar are proportional to $m_r^2$, the renormalized quadratic term in the Hamiltonian, and thus are both sensitive to the cutoff, $\Lambda$.  If $\lambda\sim {\cal O}(1)$, then either $m_r^2\gtrapprox \Lambda^2$, or it is fine-tuned to be small.  Nominally, in the standard model the cutoff is taken to be the reduced Planck scale ($\sim 10^{19} \text{ GeV}$), and thus for a 125 GeV Higgs, this represents a tuning of one part in $m_r^2/\Lambda^2 \sim 10^{34}$


Now to our solution:  As has been shown  \cite{Kibble:1978vm,Polchinski:1990py,Kaplan:2021qpv}, state-dependent terms can be easily added to the Schr\"{o}dinger equation, and thus effectively to the Hamiltonian, in a local and causal way.  We will add a term which includes the expectation value of the operator $\hat\phi(x)^2$.  

There is an important subtlety when including expectation values in the effective Hamiltonian, though it may sound trivial at first.  The expectation value is over the entire state.  This is distinct from the expectation values that appear in the equations above -- those, in effect, include a pointer \cite{Zurek:1981xq} that 'measures' the properties of the field.  We, and the rest of the universe's degrees of freedom, will be entangled with coherent states of the background Higgs expectation values.  Thus, we will distinguish the full wavefunction expectation value with double brackets -- {\it e.g.}, $\dbraket{\phi}$.

Let's start from the theory of a complex scalar above, adding a new term:
\begin{equation}\label{eq:NLHamiltonian}
    \hat H = \hat H_0 + \int d^3x \: g \,\hat\phi^\dagger\phi\dbraket{\hat\phi^\dagger\phi}  
\end{equation}
The new quartic term is in a sense a more classical contribution, as it depends on a background field. To see this, one can generate equations of motion as in \eqref{eq:LinEOM}:
\begin{eqnarray}
    \partial_t \langle \hat\phi \rangle &=& i \langle\left[ \hat H , \hat\phi \right]\rangle = \langle \hat\pi^\dagger \rangle \nonumber\\ \label{eq:NLinEOM}
    \partial^2_t\langle\hat\phi\rangle  &=& \partial_t\langle\hat\pi^\dagger\rangle =  i\langle \left[ \hat H , \hat\pi^\dagger \right]\rangle\\ &=& \nabla^2\langle\hat\phi\rangle + m_b^2 \langle\hat\phi\rangle - g \langle\hat\phi\rangle\dbraket{\hat\phi^\dagger\hat\phi}- \frac{\lambda}{2}\langle\hat\phi^\dagger\hat\phi\,\hat\phi\rangle\nonumber
\end{eqnarray}
The key point is that the commutator only picks up the $\hat\phi$'s outside of the double brackets, and thus gets the same numerical coefficient as the mass term and not the quartic.

To compute the mass of a particle, we take the effect on $\dbraket{\phi^\dagger\phi}$ to be negligible.  For example, we make the (reasonable) assumption that the particle created is done so with small probability.  In addition, when the particle is created, its momentum direction is also determined probabilistically, and thus in a typical experiment, the full wave function of the particle will be diffuse and the point of detection (for example, at the LHC, centimeters away from the production point).  Finally, it is also reasonable to assume that we (and our experiments) are a small fraction of the wavefunction of the universe.  Thus, when we expand fields around backgrounds, $\hat\phi = \phi_0 + \delta\hat\phi$, we will take, {\it e.g.}, $\dbraket{\phi} = \phi_0$.



Let's look for stable, stationary homogeneous solutions of \eqref{eq:NLinEOM}, first by making the simplifying assumption that the full wavefunction is a narrow, coherent state.  It may seems obvious that the ground state is such a state, but we need to check (as we will below) that there are no additional solutions with the wave function made up of states with widely diverse expectation values of the Higgs field.  When the wave function is made of different parts centered at different field values, we'll call this `multiple worlds'.

Stationary solutions to \eqref{eq:NLinEOM} satisfy
\begin{equation}
    0   =  m_b^2 \langle\phi\rangle - g \langle\phi\rangle\dbraket{\phi^\dagger\phi}- \frac{\lambda}{2}\langle(\phi^\dagger\phi) \,\phi\rangle
\end{equation}
Let's again break $\phi$ into real and imaginary parts, renormalize the mass, and look at solutions with $\chi = 0$:
\begin{equation}
    0 =  m_r^2 \langle\varphi\rangle - \frac{g}{2}\langle\varphi\rangle\dbraket{\varphi}^2 - \frac{\lambda}{4}\langle\varphi\rangle^3
\end{equation}
First, let's assume $\dbraket{\varphi} = \langle\varphi\rangle$ ({\it e.g.}, `one world'). We see there are stationary solutions at:
\begin{equation}
    \langle\varphi\rangle^2 = 0 \:\:{\rm and} \:\: \frac{m_r^2}{g/2 + \lambda/4}
\end{equation}
Small fluctuations around $\langle\varphi\rangle = 0$ are clearly unstable: $\partial_t^2\langle\delta\varphi\rangle \simeq  m_r^2 \langle\delta\varphi\rangle$, as $m_r^2>0$, whereas fluctuations around $\langle\varphi\rangle = \varphi_0 \equiv \sqrt{4m_r^2/(2g + \lambda)}$ are stable against  fluctuations that are a tiny part of the wave function ({\it i.e.}, fluctuations where $\dbraket{|\phi|^2}\simeq\varphi_0^2$):
\begin{eqnarray}
 \label{eq:NLfluc}
    &\simeq&  m_r^2\langle\delta\varphi\rangle - \frac{g}{2}\varphi_0^2 \langle\delta\varphi\rangle - \frac{3\lambda}{4}\varphi_0^2  \langle\delta\varphi\rangle \\ \nonumber
    &=&  - \frac{\lambda}{2}\varphi_0^2  \langle\delta\varphi\rangle
\end{eqnarray}
 From the small fluctuations in \eqref{eq:NLfluc}, we can read off the mass of a produced particle, $m_{phys}^2 = (\lambda/2)\varphi_0^2$, remarkably the same relationship as the case without the non-linear term.

One can also see that this equilibrium point is also stable under shifts of the entire central value of the coherent state (with $\dbraket{|\phi|^2}\simeq \varphi_0^2 + 2\varphi_0\langle\delta\hat\varphi\rangle$):
\begin{equation*}
\begin{split}
        \partial^2_t\langle\delta\varphi\rangle  & = m_r^2 (\varphi_0 +\langle\delta\varphi\rangle) \\& \phantom{XXX}- \frac{g}{2}(\varphi_0 + \langle\delta\varphi\rangle)^3 - \frac{\lambda}{4}(\varphi_0 + \langle\delta\varphi\rangle)^3 
    \\ & \simeq  - 2(\frac{g}{2} + \frac{\lambda}{4})\varphi_0^2  \langle\delta\varphi\rangle =  - 2 m_r^2 \langle\delta\varphi\rangle
\end{split}
\end{equation*}
thus showing that this configuration is a perturbatively stable configuration in Hilbert space.

Now let us look at a broader set of configurations which are only relevant in non-linear quantum mechanics.  Let's look for stationary states that are linear combinations of $n$ coherent states: 
\begin{equation}
   \nonumber 
|\Psi\rangle = \sum\limits_{i=1}^n \alpha_i|\phi^{\rm cl}_i\rangle
\end{equation}
such that $\langle\phi^{\rm cl}_i|\phi^{\rm cl}_j\rangle\simeq 0$ for $i\neq j$, and $\langle\phi^{\rm cl}_i| \hat{\phi}|\phi^{\rm cl}_j\rangle\simeq \delta_{ij}\phi^{\rm cl}_i $.  Thus, the static, homogeneous solutions for an observer in a sector $|\phi^{\rm cl}_i\rangle$ satisfy
\begin{equation}
    0 = m_r^2 \,\phi^{\rm cl}_i - g \,\phi^{\rm cl}_i \dbraket{\hat\phi^\dagger\hat\phi} - \frac{\lambda}{2}\phi^{\rm cl}_i|\phi^{\rm cl}_i|^2
\end{equation}
where $\dbraket{\hat\phi^\dagger\hat\phi} \simeq \sum |\alpha_i|^2|\phi^{\rm cl}_i|^2$.  Now, one can multiply this equation by $\phi^{\rm cl*}_j$, with $j\neq i$ and subtract that from the complex conjugate of the same equation with $i\leftrightarrow j$.  The first two terms cancel, leaving
\begin{equation}
    0 = \frac{\lambda}{2}\phi^{\rm cl *}_j\phi^{\rm cl}_i (|\phi^{\rm cl}_i|^2 - |\phi^{\rm cl}_j|^2)
\end{equation}
These equations have solutions with either $\phi^{\rm cl}_{i\,{\rm or}\, j}  =0$ or $\phi^{\rm cl}_i =  \phi^{\rm cl}_j e^{i \xi_j}\neq 0$ with arbitrary phase $\xi_j$.  Since this equation is true for all $i,j$, all $\phi^{\rm cl}_i$ are either zero or equal in magnitude.  Thus, $\dbraket{\hat\phi^\dagger\hat\phi} =  |\bar\alpha|^2|\phi^{\rm cl}_0|^2$, where $\bar\alpha^2=\sum |\alpha_i|^2$ and $|\phi^{\rm cl}_0|^2 \equiv |\phi^{\rm cl}_i|^2$ for any $i$ in which $\phi^{\rm cl}_i \neq0$.  There are three classes of solutions (using the notation $\langle\phi^{\rm cl}_i|\hat O|\phi^{\rm cl}_i\rangle\equiv\langle\hat O\rangle_i$:

\begin{itemize}
    \item All $\phi^{\rm cl}_i = 0$. This is clearly unstable, as $\partial_t^2\langle\delta\hat\phi\rangle_i = m_r^2\langle\delta\hat\phi\rangle_i$  

    \item  $\phi^{\rm cl}_i=0$ for $i=1,2,...,k$ with $0<k<n$, and the remaining $|\phi^{\rm cl}_j|^2 = |\phi^{\rm cl}_0|^2 =m_r^2/(g \,\bar\alpha^2 + \lambda/2)$, with $0<\bar\alpha^2<1$.  Homogeneous fluctuations around the vanishing $\langle\phi_i\rangle$ are unstable: 
        \begin{equation}
            \partial_t^2\langle\delta\hat\phi\rangle_i = m_r^2(\frac{ (\lambda/2)}{g \bar\alpha^2 + (\lambda/2)})\langle\delta\hat\phi\rangle_i
        \end{equation}

    \item Finally, if all $\phi^{\rm cl}_i\neq 0$, $\bar\alpha^2=1$ and we return to stable minima.  We expand $\hat\phi = (\hat\varphi+\hat\chi)/\sqrt{2}$ and without loss of generality take $\phi_0 =\sqrt{2}\varphi_0 \equiv \sqrt{m_r^2/(g+\lambda/2)}$ real and positive and expand in the non-flat direction, $\langle\hat\varphi\rangle_i=\varphi^{\rm cl}_0 + \langle\delta\hat\varphi\rangle_i$:
        \begin{equation}
            \begin{split}
            \partial_t^2\langle\delta\hat\varphi\rangle_i &=  \: m_r^2\langle\delta\hat\varphi\rangle_i  - \frac{g}{2} \varphi_0^2 \langle\delta\hat\varphi\rangle_i - \frac{ \lambda}{4} 3\varphi_0^2 \langle \delta\hat\varphi \rangle_i \nonumber  \\&    =- \frac{ \lambda}{2} \varphi_0^2 \langle \delta\hat\varphi\rangle_i 
            \end{split}
        \end{equation}
        This scenario is simply the one-world case -- the mass squared of fluctuations in the real direction is $\lambda\varphi_0^2/2$ and zero in the imaginary direction (the Nambu-Goldstone boson).
\end{itemize}

The general conclusion is that the only stable solutions have the same fixed magnitude for the field in all parts of the wave function, and that small fluctuations represent particles of squared mass $\lambda\varphi_0^2/2$, regardless of the existence of non-linear term.



Here is the solution to the hierarchy problem: make $g$ large.  If we assume that the renormalized squared mass $m_r^2$ is at its natural scale (the UV cutoff of the theory), we can take $g \sim m_r^2/v^2 \gg 1$, where $v$ is the standard electroweak vacuum expectation value (vev).  For a normal-sized quartic coupling, $\lambda \sim {\cal O}(1)$, the non-linear theory can naturally have $\varphi_0^2 =v^2 \ll m_r^2$, while the squared mass of an elementary particle has the standard relationship to its vev, $m_{phys}^2 = \lambda v^2/2$.

Coupling this field to gauge fields and fermions and expanding the field to an SU(2) doublet does not qualitatively change this story.  The additional interactions also renormalize the scalar mass.  However, it is the large value of $g$ that suppresses the vev of the field and mass of the physical particle -- there is no cancellation required and thus there is no restriction on the specific (state-independent) couplings of the theory.



One might be concerned that adding a large coupling $g$ would automatically generate a breakdown of a perturbative analysis of this theory.  However, $g$ is not a coupling in the usual sense as it doesn't generate particle-particle scattering.  In the limit where the regulated $\dbraket{\phi^\dagger\phi}$ is homogeneous and isotropic, $m_r^2 - g\dbraket{\phi^\dagger\phi}$ is just the squared mass parameter of particle excitations.  There are two reasons why particle excitations will have (in experimental processes to date) a negligible affect on the value of $\dbraket{\phi^\dagger\phi}$.  The first is the fact that the process of creating a particle is both weak and typically spread over a macroscopic volume.  For example, when protons collide at the large hadron collider, the wave function of colliding protons is spread over a volume of at least $L^3 \sim 1\mu{\rm m}^3$, in which case, the wave function of the Higgs created would be spread over this volume.  For a one-particle Higgs state spread over a volume $L^3$, the quantity $\langle\phi^\dagger\phi\rangle \propto 1/(L^3 m_{phys})$.  In the full wave function expectation value, this will be additionally suppressed by the probability of producing a Higgs particle in the specific proton-proton collision, which, for a micron-sized wave function of incoming protons, is roughly $\sigma_{pp\rightarrow h} /L^2 \sim 10^{-26}$ using the estimated Higgs-production cross section at the LHC, $\sigma_{pp\rightarrow h}\sim 50 $ pb \cite{Anastasiou:2016cez}).  Thus, the change to $g\dbraket{\phi^\dagger\phi}$ in creating the one-particle state for $g \simeq \Lambda^2/v^2$ is roughly $\sim \Lambda^2 \sigma_{pp\rightarrow h}/(v^3 L^5) \sim 10^{-63}\Lambda^2$.  Clearly, $\Lambda$ can be quite large without affecting particle production and the mass of the Higgs.

A second reason why particle production will likely not have an effect on the background, even for much smaller $L$ and much larger cross sections, is because it is unlikely that our part of the wave function, which we'll call $|{\rm Us}\rangle$, dominates the wave function of the universe, $|\Psi\rangle$.  If we write $|\Psi\rangle = \alpha|{\rm Us}\rangle + \beta | {\rm M}\rangle$, it is natural for $\alpha$ to be extremely small, as it is the fraction of the wave function of the universe that is inhabited by the process that locally produced the particle.  For example, if there was a period of inflation and the location of the Milky Way and the Earth were in part deterimed by a quantum fluctuation of the inflaton, then $\alpha$ will be exponentially suppressed.  Thus, generally, one can have $g\gg1$ and still have a perturbative description.  

So far, there is nothing to suggest the cutoff itself cannot be as high as the Planck scale. It would require $g\sim 10^{34}$.  While this is a very large number, as discussed above, it should not affect the perceived Higgs properties in Higgs production.  In addition, our fraction of the wave function could easily be much smaller, thus preserving a perturbative standard model in our part of the wave function.  As with technically natural theories, it would be interesting to speculate what physics above the cutoff could naturally generate such a large dimensionless number.

On the other hand, when the wave function of the universe includes more interesting (inhomogeneous) configurations, the behavior becomes very non-linear and thus one expects strong differences in the early universe from the normal weakly coupled case.  In addition, a large value of $g$ does limit the range of validity of the theory with respect to values of the field.  For field values larger than ${\cal O}(v)$ away from the minimum (in the full wave-function), the field would oscillate at a frequency $\sim\sqrt{g\dbraket{\phi^2}} > \Lambda$ and thus could decay into particles of mass higher than the cutoff.  Thus we restrict states to $\dbraket{\phi^2}\sim v^2$, though in principle $\langle\phi\rangle$ could be much larger in a tiny part of the wave function without exciting heavy states.  It would be interesting to implement a complete model of the cutoff ({\it e.g.}, supersymmetry) and see that the theory remains self-consistent.



A solution to the hierarchy problem should include robust insensitivity to initial conditions.  Normally, this is simply a requirement that the universe naturally ends up in the correct vacuum.  Here, while there are no undesirable vacua, one can ask what it takes to relax the field value to its natural minimum.  It is easy to show that it is in fact a robust outcome in some scenarios, and an interesting point of exploration in others.

The simplest scenario is one in which there is a period of inflation that ends with reheating the universe to temperatures below the weak scale.  Assuming the Higgs sector does not dominate the energy of the initial state, the gravitational physics can be included in the equations of motion of the Higgs as a classical background:
\begin{equation}
\begin{split}
    \partial_t^2 \langle\hat\phi\rangle + 3{\cal H}\partial_t\langle\hat\phi\rangle = & \frac{1}{a^2(t)}\nabla^2\langle\hat\phi\rangle  + m_r^2\langle\hat\phi\rangle \\& - g\langle\hat\phi\rangle \dbraket{\hat\phi^\dagger\hat\phi} - \frac{\lambda}{2} \langle\hat\phi\hat\phi^\dagger\hat\phi\rangle
\end{split}
\end{equation}
where we've constructed the Hamiltonian with the standard FRW metric, $ds^2=dt^2 - a^2(t)d\mathbf{x}^2$ and ${\cal H}\equiv \dot{a}/a$ is the Hubble rate.  The friction term, as in the standard case, drives the expectation value of $\hat\phi$ to its minimum.  During a period of inflation, one would expect the Higgs to rapidly descend to its minimum value with tiny width.

One can also check that a homogeneous coherent state with an initial value of $|\langle\hat\phi\rangle| = \phi_{\rm in} \neq\phi_0$ will oscillate rapidly -- initially at frequency $\Omega\sim g\phi_{\rm in}^2$ when the quartic term dominates -- and decay rapidly into higher momentum modes in the standard language of parametric resonance \cite{Kofman:1994rk}.  Essentially, $\langle\phi\rangle$ (and the coherent oscillations) will decrease while $\langle\phi^\dagger\phi\rangle$ will increase until every part of the wavefunction looks like a thermal state.  One can in fact show that there is an energy functional, $\langle\hat H_0\rangle + (1/2)g\int d^3x\langle|\hat\phi|^2\rangle\langle|\hat\phi|^2\rangle$, which is conserved in Minkowski space and reduces in an expanding universe.

A more interesting possibility is the case where the Higgs is a non-trivial part of the energy density of the universe.  In this case, different parts of the wave function will have different metrics, and thus the classical gravitational background is no more.  This would require an analysis beyond the scope of this paper. 


There are other similar models with differing non-linear terms that would also solve the hierarchy problem.  For example, instead of the $g$ term in \eqref{eq:NLHamiltonian}, we could have added $\dbraket{\hat\phi^\dagger}\dbraket{\hat\phi}$.  This term has expectation values of operators which transform under the gauge symmetry, and yet is naively gauge invariant.  This version may have interesting cosmological differences, but otherwise has mostly similar vacua as the original term.  A more exotic term would be $ \mu^4 \hat{\phi}^\dagger\hat\phi/\dbraket{\hat\phi^\dagger\hat\phi}$, which would allow large values of $\langle\phi\rangle$ in the effective field theory, but not values near the origin.  This term has some preferred vacua, but allows other solutions where the hierarchy problem is not solved.  Finally, terms like $\hat\phi^\dagger\hat\phi\hat\phi^\dagger\dbraket{\hat\phi}$ themselves do not solve the hierarchy problem, but could be added to \eqref{eq:NLHamiltonian} with small coefficients.  Such a term is again naively gauge invariant, but would change the Higgs interactions relative to the standard model, and thus potentially testable in future colliders.  It is thus worth checking whether such terms predict strong coupling/breakdown of unitarity at high energies.




While the simplest model has (as far as we can tell) a dearth of new signals at colliders, it could have profoundly different cosmological implications.  One that stands out is the nature of the transition from a hot universe to an electroweak symmetry breaking vacuum.  While there may be a first-order phase transition hiding in some initial conditions, there is also the potential of rapid, out-of-equilibrium particle production.  This process may help with new types of models of electroweak baryogenesis and should be explored.

\section*{Acknowledgements}
We thank Gian Giudice, Riccardo Rattazzi, Raman Sundrum, and especially Anson Hook for fruitful discussions. This work was supported by the U.S.~Department of Energy~(DOE), Office of Science, National Quantum Information Science Research Centers, Superconducting Quantum Materials and Systems Center~(SQMS) under Contract No.~DE-AC02-07CH11359. D.K.\ and S.R.\ are supported in part by the U.S.~National Science Foundation~(NSF) under Grant No.~PHY-1818899.  D.K.\ is also supported by the Simons Investigator Award no.~144924.
S.R.\ is also supported by the~DOE under a QuantISED grant for MAGIS.
S.R.\ is also supported by the Simons Investigator Award No.~827042. 
\bibliographystyle{apsrev4-2}
\bibliography{refs}

\begin{thebibliography}{22}%
\makeatletter
\providecommand \@ifxundefined [1]{%
 \@ifx{#1\undefined}
}%
\providecommand \@ifnum [1]{%
 \ifnum #1\expandafter \@firstoftwo
 \else \expandafter \@secondoftwo
 \fi
}%
\providecommand \@ifx [1]{%
 \ifx #1\expandafter \@firstoftwo
 \else \expandafter \@secondoftwo
 \fi
}%
\providecommand \natexlab [1]{#1}%
\providecommand \enquote  [1]{``#1''}%
\providecommand \bibnamefont  [1]{#1}%
\providecommand \bibfnamefont [1]{#1}%
\providecommand \citenamefont [1]{#1}%
\providecommand \href@noop [0]{\@secondoftwo}%
\providecommand \href [0]{\begingroup \@sanitize@url \@href}%
\providecommand \@href[1]{\@@startlink{#1}\@@href}%
\providecommand \@@href[1]{\endgroup#1\@@endlink}%
\providecommand \@sanitize@url [0]{\catcode `\\12\catcode `\$12\catcode `\&12\catcode `\#12\catcode `\^12\catcode `\_12\catcode `\%12\relax}%
\providecommand \@@startlink[1]{}%
\providecommand \@@endlink[0]{}%
\providecommand \url  [0]{\begingroup\@sanitize@url \@url }%
\providecommand \@url [1]{\endgroup\@href {#1}{\urlprefix }}%
\providecommand \urlprefix  [0]{URL }%
\providecommand \Eprint [0]{\href }%
\providecommand \doibase [0]{https://doi.org/}%
\providecommand \selectlanguage [0]{\@gobble}%
\providecommand \bibinfo  [0]{\@secondoftwo}%
\providecommand \bibfield  [0]{\@secondoftwo}%
\providecommand \translation [1]{[#1]}%
\providecommand \BibitemOpen [0]{}%
\providecommand \bibitemStop [0]{}%
\providecommand \bibitemNoStop [0]{.\EOS\space}%
\providecommand \EOS [0]{\spacefactor3000\relax}%
\providecommand \BibitemShut  [1]{\csname bibitem#1\endcsname}%
\let\auto@bib@innerbib\@empty
\bibitem [{Note1()}]{Note1}%
  \BibitemOpen
  \bibinfo {note} {A first discussion appears in print in \cite {PhysRevD.20.2619}, but Susskind (and many others) credit Ken Wilson for the original insight of a naturalness problem. It is more thoroughly discussed in 't Hooft's Cargese lectures \cite {tHooft:1980xss}.}\BibitemShut {Stop}%
\bibitem [{\citenamefont {Dimopoulos}\ and\ \citenamefont {Georgi}(1981)}]{Dimopoulos:1981zb}%
  \BibitemOpen
  \bibfield  {author} {\bibinfo {author} {\bibfnamefont {S.}~\bibnamefont {Dimopoulos}}\ and\ \bibinfo {author} {\bibfnamefont {H.}~\bibnamefont {Georgi}},\ }\href {https://doi.org/10.1016/0550-3213(81)90522-8} {\bibfield  {journal} {\bibinfo  {journal} {Nucl. Phys. B}\ }\textbf {\bibinfo {volume} {193}},\ \bibinfo {pages} {150} (\bibinfo {year} {1981})}\BibitemShut {NoStop}%
\bibitem [{\citenamefont {Kaplan}\ and\ \citenamefont {Georgi}(1984)}]{Kaplan:1983fs}%
  \BibitemOpen
  \bibfield  {author} {\bibinfo {author} {\bibfnamefont {D.~B.}\ \bibnamefont {Kaplan}}\ and\ \bibinfo {author} {\bibfnamefont {H.}~\bibnamefont {Georgi}},\ }\href {https://doi.org/10.1016/0370-2693(84)91177-8} {\bibfield  {journal} {\bibinfo  {journal} {Phys. Lett. B}\ }\textbf {\bibinfo {volume} {136}},\ \bibinfo {pages} {183} (\bibinfo {year} {1984})}\BibitemShut {NoStop}%
\bibitem [{\citenamefont {Arkani-Hamed}\ \emph {et~al.}(1998)\citenamefont {Arkani-Hamed}, \citenamefont {Dimopoulos},\ and\ \citenamefont {Dvali}}]{Arkani-Hamed:1998jmv}%
  \BibitemOpen
  \bibfield  {author} {\bibinfo {author} {\bibfnamefont {N.}~\bibnamefont {Arkani-Hamed}}, \bibinfo {author} {\bibfnamefont {S.}~\bibnamefont {Dimopoulos}},\ and\ \bibinfo {author} {\bibfnamefont {G.~R.}\ \bibnamefont {Dvali}},\ }\href {https://doi.org/10.1016/S0370-2693(98)00466-3} {\bibfield  {journal} {\bibinfo  {journal} {Phys. Lett. B}\ }\textbf {\bibinfo {volume} {429}},\ \bibinfo {pages} {263} (\bibinfo {year} {1998})},\ \Eprint {https://arxiv.org/abs/hep-ph/9803315} {arXiv:hep-ph/9803315} \BibitemShut {NoStop}%
\bibitem [{\citenamefont {Navas}\ \emph {et~al.}(2024)\citenamefont {Navas} \emph {et~al.}}]{ParticleDataGroup:2024cfk}%
  \BibitemOpen
  \bibfield  {author} {\bibinfo {author} {\bibfnamefont {S.}~\bibnamefont {Navas}} \emph {et~al.} (\bibinfo {collaboration} {Particle Data Group}),\ }\href {https://doi.org/10.1103/PhysRevD.110.030001} {\bibfield  {journal} {\bibinfo  {journal} {Phys. Rev. D}\ }\textbf {\bibinfo {volume} {110}},\ \bibinfo {pages} {030001} (\bibinfo {year} {2024})}\BibitemShut {NoStop}%
\bibitem [{\citenamefont {Glioti}\ \emph {et~al.}(2025)\citenamefont {Glioti}, \citenamefont {Rattazzi}, \citenamefont {Ricci},\ and\ \citenamefont {Vecchi}}]{Glioti:2024hye}%
  \BibitemOpen
  \bibfield  {author} {\bibinfo {author} {\bibfnamefont {A.}~\bibnamefont {Glioti}}, \bibinfo {author} {\bibfnamefont {R.}~\bibnamefont {Rattazzi}}, \bibinfo {author} {\bibfnamefont {L.}~\bibnamefont {Ricci}},\ and\ \bibinfo {author} {\bibfnamefont {L.}~\bibnamefont {Vecchi}},\ }\href {https://doi.org/10.21468/SciPostPhys.18.6.201} {\bibfield  {journal} {\bibinfo  {journal} {SciPost Phys.}\ }\textbf {\bibinfo {volume} {18}},\ \bibinfo {pages} {201} (\bibinfo {year} {2025})},\ \Eprint {https://arxiv.org/abs/2402.09503} {arXiv:2402.09503 [hep-ph]} \BibitemShut {NoStop}%
\bibitem [{\citenamefont {Weinberg}(1989)}]{Weinberg:1988cp}%
  \BibitemOpen
  \bibfield  {author} {\bibinfo {author} {\bibfnamefont {S.}~\bibnamefont {Weinberg}},\ }\href {https://doi.org/10.1103/RevModPhys.61.1} {\bibfield  {journal} {\bibinfo  {journal} {Rev. Mod. Phys.}\ }\textbf {\bibinfo {volume} {61}},\ \bibinfo {pages} {1} (\bibinfo {year} {1989})}\BibitemShut {NoStop}%
\bibitem [{\citenamefont {Graham}\ \emph {et~al.}(2015)\citenamefont {Graham}, \citenamefont {Kaplan},\ and\ \citenamefont {Rajendran}}]{Graham:2015cka}%
  \BibitemOpen
  \bibfield  {author} {\bibinfo {author} {\bibfnamefont {P.~W.}\ \bibnamefont {Graham}}, \bibinfo {author} {\bibfnamefont {D.~E.}\ \bibnamefont {Kaplan}},\ and\ \bibinfo {author} {\bibfnamefont {S.}~\bibnamefont {Rajendran}},\ }\href {https://doi.org/10.1103/PhysRevLett.115.221801} {\bibfield  {journal} {\bibinfo  {journal} {Phys. Rev. Lett.}\ }\textbf {\bibinfo {volume} {115}},\ \bibinfo {pages} {221801} (\bibinfo {year} {2015})},\ \Eprint {https://arxiv.org/abs/1504.07551} {arXiv:1504.07551 [hep-ph]} \BibitemShut {NoStop}%
\bibitem [{\citenamefont {Arkani-Hamed}\ and\ \citenamefont {Dimopoulos}(2005)}]{Arkani-Hamed:2004ymt}%
  \BibitemOpen
  \bibfield  {author} {\bibinfo {author} {\bibfnamefont {N.}~\bibnamefont {Arkani-Hamed}}\ and\ \bibinfo {author} {\bibfnamefont {S.}~\bibnamefont {Dimopoulos}},\ }\href {https://doi.org/10.1088/1126-6708/2005/06/073} {\bibfield  {journal} {\bibinfo  {journal} {JHEP}\ }\textbf {\bibinfo {volume} {06}},\ \bibinfo {pages} {073}},\ \Eprint {https://arxiv.org/abs/hep-th/0405159} {arXiv:hep-th/0405159} \BibitemShut {NoStop}%
\bibitem [{\citenamefont {Dvali}(2010)}]{Dvali:2007hz}%
  \BibitemOpen
  \bibfield  {author} {\bibinfo {author} {\bibfnamefont {G.}~\bibnamefont {Dvali}},\ }\href {https://doi.org/10.1002/prop.201000009} {\bibfield  {journal} {\bibinfo  {journal} {Fortsch. Phys.}\ }\textbf {\bibinfo {volume} {58}},\ \bibinfo {pages} {528} (\bibinfo {year} {2010})},\ \Eprint {https://arxiv.org/abs/0706.2050} {arXiv:0706.2050 [hep-th]} \BibitemShut {NoStop}%
\bibitem [{\citenamefont {Arkani-Hamed}\ \emph {et~al.}(2016)\citenamefont {Arkani-Hamed}, \citenamefont {Cohen}, \citenamefont {D'Agnolo}, \citenamefont {Hook}, \citenamefont {Kim},\ and\ \citenamefont {Pinner}}]{Arkani-Hamed:2016rle}%
  \BibitemOpen
  \bibfield  {author} {\bibinfo {author} {\bibfnamefont {N.}~\bibnamefont {Arkani-Hamed}}, \bibinfo {author} {\bibfnamefont {T.}~\bibnamefont {Cohen}}, \bibinfo {author} {\bibfnamefont {R.~T.}\ \bibnamefont {D'Agnolo}}, \bibinfo {author} {\bibfnamefont {A.}~\bibnamefont {Hook}}, \bibinfo {author} {\bibfnamefont {H.~D.}\ \bibnamefont {Kim}},\ and\ \bibinfo {author} {\bibfnamefont {D.}~\bibnamefont {Pinner}},\ }\href {https://doi.org/10.1103/PhysRevLett.117.251801} {\bibfield  {journal} {\bibinfo  {journal} {Phys. Rev. Lett.}\ }\textbf {\bibinfo {volume} {117}},\ \bibinfo {pages} {251801} (\bibinfo {year} {2016})},\ \Eprint {https://arxiv.org/abs/1607.06821} {arXiv:1607.06821 [hep-ph]} \BibitemShut {NoStop}%
\bibitem [{Note2()}]{Note2}%
  \BibitemOpen
  \bibinfo {note} {The basic idea discussed here can also potentially be applied to get around the Weinberg no-go theorem for the cosmological constant. We leave this for future work.}\BibitemShut {Stop}%
\bibitem [{\citenamefont {Kibble}(1978)}]{Kibble:1978vm}%
  \BibitemOpen
  \bibfield  {author} {\bibinfo {author} {\bibfnamefont {T.~W.~B.}\ \bibnamefont {Kibble}},\ }\href {https://doi.org/10.1007/BF01940762} {\bibfield  {journal} {\bibinfo  {journal} {Commun. Math. Phys.}\ }\textbf {\bibinfo {volume} {64}},\ \bibinfo {pages} {73} (\bibinfo {year} {1978})}\BibitemShut {NoStop}%
\bibitem [{\citenamefont {Kaplan}\ and\ \citenamefont {Rajendran}(2022)}]{Kaplan:2021qpv}%
  \BibitemOpen
  \bibfield  {author} {\bibinfo {author} {\bibfnamefont {D.~E.}\ \bibnamefont {Kaplan}}\ and\ \bibinfo {author} {\bibfnamefont {S.}~\bibnamefont {Rajendran}},\ }\href {https://doi.org/10.1103/PhysRevD.105.055002} {\bibfield  {journal} {\bibinfo  {journal} {Phys. Rev. D}\ }\textbf {\bibinfo {volume} {105}},\ \bibinfo {pages} {055002} (\bibinfo {year} {2022})},\ \Eprint {https://arxiv.org/abs/2106.10576} {arXiv:2106.10576 [hep-th]} \BibitemShut {NoStop}%
\bibitem [{\citenamefont {Kibble}(1979)}]{Kibble:1978tm}%
  \BibitemOpen
  \bibfield  {author} {\bibinfo {author} {\bibfnamefont {T.~W.~B.}\ \bibnamefont {Kibble}},\ }\href {https://doi.org/10.1007/BF01225149} {\bibfield  {journal} {\bibinfo  {journal} {Commun. Math. Phys.}\ }\textbf {\bibinfo {volume} {65}},\ \bibinfo {pages} {189} (\bibinfo {year} {1979})}\BibitemShut {NoStop}%
\bibitem [{Note3()}]{Note3}%
  \BibitemOpen
  \bibinfo {note} {The remaining terms include wave-function renormalization and finite state dependent terms which will not affects the hierarchy problem in a qualitative way.}\BibitemShut {Stop}%
\bibitem [{\citenamefont {Polchinski}(1991)}]{Polchinski:1990py}%
  \BibitemOpen
  \bibfield  {author} {\bibinfo {author} {\bibfnamefont {J.}~\bibnamefont {Polchinski}},\ }\href {https://doi.org/10.1103/PhysRevLett.66.397} {\bibfield  {journal} {\bibinfo  {journal} {Phys. Rev. Lett.}\ }\textbf {\bibinfo {volume} {66}},\ \bibinfo {pages} {397} (\bibinfo {year} {1991})}\BibitemShut {NoStop}%
\bibitem [{\citenamefont {Zurek}(1981)}]{Zurek:1981xq}%
  \BibitemOpen
  \bibfield  {author} {\bibinfo {author} {\bibfnamefont {W.~H.}\ \bibnamefont {Zurek}},\ }\href {https://doi.org/10.1103/PhysRevD.24.1516} {\bibfield  {journal} {\bibinfo  {journal} {Phys. Rev. D}\ }\textbf {\bibinfo {volume} {24}},\ \bibinfo {pages} {1516} (\bibinfo {year} {1981})}\BibitemShut {NoStop}%
\bibitem [{\citenamefont {Anastasiou}\ \emph {et~al.}(2016)\citenamefont {Anastasiou}, \citenamefont {Duhr}, \citenamefont {Dulat}, \citenamefont {Furlan}, \citenamefont {Gehrmann}, \citenamefont {Herzog}, \citenamefont {Lazopoulos},\ and\ \citenamefont {Mistlberger}}]{Anastasiou:2016cez}%
  \BibitemOpen
  \bibfield  {author} {\bibinfo {author} {\bibfnamefont {C.}~\bibnamefont {Anastasiou}}, \bibinfo {author} {\bibfnamefont {C.}~\bibnamefont {Duhr}}, \bibinfo {author} {\bibfnamefont {F.}~\bibnamefont {Dulat}}, \bibinfo {author} {\bibfnamefont {E.}~\bibnamefont {Furlan}}, \bibinfo {author} {\bibfnamefont {T.}~\bibnamefont {Gehrmann}}, \bibinfo {author} {\bibfnamefont {F.}~\bibnamefont {Herzog}}, \bibinfo {author} {\bibfnamefont {A.}~\bibnamefont {Lazopoulos}},\ and\ \bibinfo {author} {\bibfnamefont {B.}~\bibnamefont {Mistlberger}},\ }\href {https://doi.org/10.1007/JHEP05(2016)058} {\bibfield  {journal} {\bibinfo  {journal} {JHEP}\ }\textbf {\bibinfo {volume} {05}},\ \bibinfo {pages} {058}},\ \Eprint {https://arxiv.org/abs/1602.00695} {arXiv:1602.00695 [hep-ph]} \BibitemShut {NoStop}%
\bibitem [{\citenamefont {Kofman}\ \emph {et~al.}(1994)\citenamefont {Kofman}, \citenamefont {Linde},\ and\ \citenamefont {Starobinsky}}]{Kofman:1994rk}%
  \BibitemOpen
  \bibfield  {author} {\bibinfo {author} {\bibfnamefont {L.}~\bibnamefont {Kofman}}, \bibinfo {author} {\bibfnamefont {A.~D.}\ \bibnamefont {Linde}},\ and\ \bibinfo {author} {\bibfnamefont {A.~A.}\ \bibnamefont {Starobinsky}},\ }\href {https://doi.org/10.1103/PhysRevLett.73.3195} {\bibfield  {journal} {\bibinfo  {journal} {Phys. Rev. Lett.}\ }\textbf {\bibinfo {volume} {73}},\ \bibinfo {pages} {3195} (\bibinfo {year} {1994})},\ \Eprint {https://arxiv.org/abs/hep-th/9405187} {arXiv:hep-th/9405187} \BibitemShut {NoStop}%
\bibitem [{\citenamefont {Susskind}(1979)}]{PhysRevD.20.2619}%
  \BibitemOpen
  \bibfield  {author} {\bibinfo {author} {\bibfnamefont {L.}~\bibnamefont {Susskind}},\ }\href {https://doi.org/10.1103/PhysRevD.20.2619} {\bibfield  {journal} {\bibinfo  {journal} {Phys. Rev. D}\ }\textbf {\bibinfo {volume} {20}},\ \bibinfo {pages} {2619} (\bibinfo {year} {1979})}\BibitemShut {NoStop}%
\bibitem [{\citenamefont {'t~Hooft}\ \emph {et~al.}(1980)\citenamefont {'t~Hooft}, \citenamefont {Itzykson}, \citenamefont {Jaffe}, \citenamefont {Lehmann}, \citenamefont {Mitter}, \citenamefont {Singer},\ and\ \citenamefont {Stora}}]{tHooft:1980xss}%
  \BibitemOpen
  \bibinfo {editor} {\bibfnamefont {G.}~\bibnamefont {'t~Hooft}}, \bibinfo {editor} {\bibfnamefont {C.}~\bibnamefont {Itzykson}}, \bibinfo {editor} {\bibfnamefont {A.}~\bibnamefont {Jaffe}}, \bibinfo {editor} {\bibfnamefont {H.}~\bibnamefont {Lehmann}}, \bibinfo {editor} {\bibfnamefont {P.~K.}\ \bibnamefont {Mitter}}, \bibinfo {editor} {\bibfnamefont {I.~M.}\ \bibnamefont {Singer}},\ and\ \bibinfo {editor} {\bibfnamefont {R.}~\bibnamefont {Stora}},\ eds.,\ \href {https://doi.org/10.1007/978-1-4684-7571-5} {\emph {\bibinfo {title} {{Recent Developments in Gauge Theories. Proceedings, Nato Advanced Study Institute, Cargese, France, August 26 - September 8, 1979}}}},\ Vol.~\bibinfo {volume} {59}\ (\bibinfo {year} {1980})\BibitemShut {NoStop}%
\end{thebibliography}%

\end{document}